\documentclass[aps,floatfix,showpacs,showkeys,amsmath,amssymb]{revtex4}
\usepackage[T1]{fontenc}
\usepackage[latin1]{inputenc}
\usepackage{graphicx}
\usepackage{dcolumn}
\usepackage{bm}

\newcommand{\be}{\begin{equation}}
\newcommand{\ee}{\end{equation}}
\newcommand{\bea}{\begin{eqnarray}}
\newcommand{\eea}{\end{eqnarray}}

\newcommand{\gz}{G_{HF}}

\newcommand{\grpa}{G^{(\alpha)}_{RPA}}

\newcommand{\grpaz}{G^{(0)}_{RPA}}
\newcommand{\grpaum}{G^{(1,M_S)}_{RPA}}
\newcommand{\grpaump}{G^{(1,M_S^\prime)}_{RPA}}

\newcommand{\romq}{\mathrm q}
\newcommand{\romk}{\mathrm k}
\newcommand{\romkf}{\mathrm k_F}

\begin{document}

\title{Effects of spin-orbit interaction on nuclear response and neutrino mean free path}

\author{J. Margueron}
\affiliation{Institut de Physique Nucl\'eaire, Universit\'e Paris
Sud F-91406 Orsay CEDEX, France}
\author{J. Navarro}
\affiliation{IFIC (CSIC -
Universidad de Valencia), Apdo. 22085, E-46.071-Valencia, Spain}
\author{N. Van Giai}
\affiliation{Institut de Physique Nucl\'eaire, Universit\'e Paris
Sud F-91406 Orsay Cedex, France}

\date{ \today}

\begin{abstract}
The effects of the spin-orbit component of the particle-hole
interaction on nuclear response functions and neutrino mean free
path are examined. A complete treatment of the full Skyrme
interaction in the case of symmetric nuclear matter and pure neutron
matter is given. Numerical results for neutron matter are discussed.
It is shown that the effects of the spin-orbit interaction remain
small, even at momentum transfer larger than the Fermi momentum. The
neutrino mean free paths are marginally affected.
\end{abstract}



\maketitle

\section{Introduction}
The mean field theory of nuclear systems is a well developed tool
for the microscopic study of finite nuclei as well as infinite
matter. Many effective two-body interactions have been determined
with the goal of achieving a self-consistent mean field description of
nuclear ground states and excited states through the Hartree-Fock
(HF) and random phase approximation (RPA) approaches. The
Skyrme-type interactions are among the most frequently used
effective forces, and this is due to the fact that they are
relatively simple and yet, they can give a fairly accurate
description of finite nuclei data\cite{vau72,bender-reinhardt}.

Homogeneous infinite matter is just an idealized object but it can
be a very useful testing ground for various theories because
practical computations are easier to carry out than in finite
nuclear systems. It is also a good representation of the internal
regions of stellar objects such as neutron stars. The matter
equation of state and the response of matter to various external
probes are important physical properties. For instance, neutrino
mean free paths can be deduced from nuclear response
functions\cite{iwa,reddy}. Therefore, HF and RPA studies of infinite
matter have always been carried out along with finite nuclei
studies. The formalism for computing RPA response functions in
infinite Fermion systems with Skyrme-type interactions has been
shown in Ref.\cite{gar92} where applications were made for symmetric
nuclear matter while calculations for neutron matter have been
performed in Ref.\cite{nav99}. The two-body spin-orbit component has
been ignored in the previous studies of the nuclear response
function. Whereas it is true that the spin-orbit term does not
contribute to quantities as the equation of state of saturated spin
systems, one should consider also situations where an external
operator can induce spin oscillations which can manifest in the
response function. The aim of this work is to give a complete
treatment of the full Skyrme interaction in the case of symmetric
nuclear matter and pure neutron matter. The spin-orbit component of
the particle-hole (p-h) interaction couples the spin channels in the
response function. However, it will be shown that the effect remains
small, even at momentum transfer larger than the Fermi momentum,
while the neutrino mean free paths are marginally affected.

The outline of the paper is as follows. In Sec.~II we present the
general method for calculating RPA response functions with the full
Skyrme interaction, generalizing the method of Garcia-Recio et
al.\cite{gar92} to the case of the spin-orbit interaction. In
Sec.~III the results obtained for response functions in symmetric
nuclear matter and pure neutron matter, as well as neutrino mean
free paths are discussed. Concluding remarks are given in Sec.~IV.

\section{Formalism}
\label{formalism}
\subsection{Definitions}
\begin{figure}[h]
\includegraphics[scale=0.25,clip]{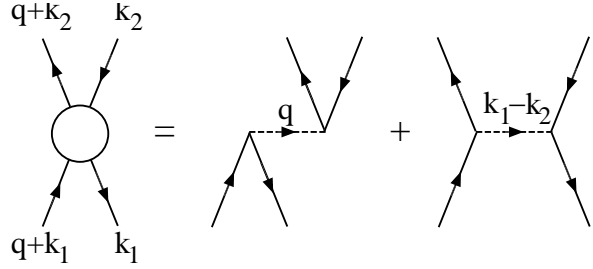}
\caption{Direct and exchange parts of the ph interaction.}
\label{qk1k2}
\end{figure}

A general two-body interaction in momentum representation depends at
most on 4 momenta. Because of momentum conservation there are
actually 3 independent momenta. For the p-h case we choose these
independent variables to be the initial (final) momentum ${\bf k_1}$
(${\bf k_2}$) of the hole and the external momentum transfer ${\bf
q}$. This is illustrated by Fig.~\ref{qk1k2}. We will denote by
$\alpha = (S,M;I,Q)$ the spin and isospin p-h channels with $S$=0
(1) for the non spin-flip (spin-flip) channel, $I$=0 (1) the
isoscalar (isovector) channel, $M$ and $Q$ being the third
components of $S$ and $I$.

Let us consider an infinite nuclear medium at zero temperature and
unpolarized both in spin and isospin spaces. At mean field level
this system is described as an ensemble of independent nucleons
moving in a self-consistent mean field generated by the starting
effective interaction treated in the Hartree-Fock (HF)
approximation. The momentum dependent HF mean field, or self-energy
determines the single-particle spectrum $\epsilon(\romk)$ and the
Fermi level $\epsilon(\romkf)$.

To calculate the response of the medium to an external field it is
convenient to introduce the Green's function, or retarded p-h
propagator $G^{(\alpha)}({\bf q},\omega,{\bf k}_1)$. From now on we
choose the $z$ axis along the direction of ${\bf q}$. In the HF
approximation, the p-h Green's function is the free retarded p-h
propagator~\cite{fet71}:
\be G_{HF}(\romq,\omega,{\bf k}_1) = \frac{f({\romk}_1)-f(\vert{\bf
k}_1+{\bf q}\vert)} {\omega + \epsilon({\romk}_1) -
\epsilon(\vert{\bf q}+{\bf k}_1)\vert + i \eta}~, \label{GHF} \ee
where the Fermi-Dirac distribution $f$ is defined for a given
temperature $T$ and chemical potential $\mu$ as
$f(\romk)=[1+e^{(e(\romk)-\mu)/T}]^{-1}$. The HF Green's function
$G_{HF}$ does not depend on the spin-isospin channel $\alpha$. To go
beyond the HF mean field approximation one takes into account the
long-range type of correlations by  resumming a class of p-h
diagrams and one obtains the well-known random phase
approximation~\cite{fet71}. The interaction appearing in the RPA is
the p-h residual interaction whose matrix element
including exchange can be written as:
\be V_{ph}^{(\alpha,\alpha')}(q,{\bf k}_1,{\bf k}_2) \equiv \langle
{\bf q}+{\bf k}_1, {\bf k}_1^{-1},(\alpha) | V | {\bf q}+{\bf k}_2,
{\bf k}_2^{-1} (\alpha') \rangle \label{Vph}~. \ee
The RPA correlated Green's function
$G_{RPA}^{(\alpha)}(\romq,\omega,{\bf k}_1)$ satisfies the
Bethe-Salpeter equation:
 \be G_{RPA}^{(\alpha)}(\romq,\omega,{\bf
k}_1) = G_{HF}(\romq,\omega,{\bf k}_1) + G_{HF}(\romq,\omega,{\bf
k}_1) \, \sum_{(\alpha')} \, \int \frac{{\rm d}^3k_2}{(2 \pi)^3}
V_{ph}^{(\alpha,\alpha')}(\romq,{\bf k}_1,{\bf k}_2)
G_{RPA}^{(\alpha')}(\romq,\omega,{\bf k}_2)~. \label{eqBS} \ee
Finally, the response function $\chi^{(\alpha)}(\romq,\omega)$ in
the infinite medium is related to the p-h Green's function by:
\be \chi_{RPA}^{(\alpha)}(\romq,\omega) = g \int \frac{{\rm d}^3
k_1}{(2 \pi)^3} G_{RPA}^{(\alpha)}(\romq,\omega,{\bf k}_1)~,
\label{chi} \ee
where the spin-isospin degeneracy factor $g$ is 4
for symmetric nuclear matter and 2 for pure neutron matter. The
Lindhard function $\chi_{HF}$ is obtained when the free p-h
propagator $G_{HF}$ is used in Eq.~(\ref{chi}).

In the following parts of this work we will often deal with
integrals similar to those appearing in Eqs.~(\ref{eqBS}-\ref{chi})
and we will adopt the notation $\langle
V_{ph}^{(\alpha\alpha')}G_{RPA}^{(\alpha')}\rangle$, $\langle
G_{RPA}^{(\alpha)}\rangle$ for such quantities.

\subsection{The p-h interaction}

The central component of the p-h interaction can be written in the
general form:
\bea V_{ph}^{(\alpha,\alpha')}(q,{\bf k}_1,{\bf k}_2) &=&
\delta(\alpha,\alpha')    \left\{ W_1^{(\alpha)} + W_2^{(\alpha)} [
\romk_1^2+\romk_2^2] - 2 W_2^{(\alpha)} \frac{4 \pi}{3} \romk_1
\romk_2 \sum_{\mu} Y^*_{1 \mu}(\hat{k_1}) Y_{1 \mu}(\hat{k_2})
\right\} \, , \label{vph-no} \eea
where the $W^{(\alpha)}_i$ are combinations of the Skyrme parameters
($t_i,x_i$) and of the transferred momentum
$\romq$.
Their detailed expressions are given in Refs.\cite{gar92} and
\cite{nav99} for the symmetric nuclear matter and pure neutron
cases, respectively. One can note that there is no coupling between
the different spin and isospin channels. The general case of matter
with an arbitrary neutron-to-proton ratio has been studied in
Ref.\cite{her97}.

The Skyrme interactions also contain a zero-range spin-orbit
term~\cite{vau72}.
It has the form $iW_{so} (\sigma_1 + \sigma_2) \cdot[{\bf k'} \times
\delta(r_{12}) {\bf k}]$, where $\sigma_i$ is the spin operator of
particle $i$, and ${\bf k}$ and ${\bf k}'$ are the relative momentum
operator of the particles acting to the right and left,
respectively. To calculate the contribution of this term to the p-h
interaction one has to evaluate the matrix element (\ref{Vph}) of
this spin-orbit interaction. As this term is density-independent
there is no rearrangement contribution and the result is just adding
the following term
\bea - \delta(I,I') \, w(I) \, \sqrt{\frac{4 \pi}{3}}
\, \romq \, W_{so} \bigg\{  & & \delta(S,1) \delta(S',0) M_S
\left[ \romk_1 Y_{1-M_S}(\hat{k_1}) - \romk_2 Y_{1-M_S}(\hat{k_2}) \right] \nonumber \\
&& + \delta(S,0) \delta(S',1) M_S^\prime \left[ \romk_1
Y_{1M_S^\prime}(\hat{k_1}) - \romk_2 Y_{1M_S^\prime}(\hat{k_2})
\right] \bigg\}~, \label{vph-so} \eea
to Eq.~(\ref{vph-no}). We have defined $ w(I) = 2 + (-1)^I $ in the
case of symmetric nuclear matter, and $w(I)=2$ for pure neutron
matter. The effect of the spin-orbit component is to couple both
$S=0$ and 1 channels.

\subsection{ Response function }

To obtain the RPA response function of Eq.~(\ref{chi}) one has to
calculate
the correlated Green's function $G_{RPA}^{(\alpha)}$. The technical
details are given in the Appendix. The response function can then be
written in the form:
\bea \frac{\chi_{HF}}{ \chi^{(\alpha)}_{RPA}} &=& 1-\widetilde
{W}_1^{(\alpha)}\chi_{0} - 2 W_2^{(\alpha)}
\left[\frac{\romq^2}{4}-\left(\frac{\omega m^*}{\romq}\right)^2
\frac{1}{1-\frac{\displaystyle m^*\romk_F^3}{\displaystyle 3\pi^2}
W_2^{(\alpha)}}\right] \chi_{0} \nonumber \\ &
&+2W_2^{(\alpha)}\left(\frac{\romq^2}{2}\chi_{0}-\romk^2_F\chi_2
\right) +[W_2^{(\alpha)}\romk^2_F]^2
\left[\chi_2^2-\chi_{0}\chi_4+\left(\frac{\omega
m^*}{\romk_F^2}\right)^2\chi_{0}^2 -
\frac{m^*}{6\pi^2\romk_F}\romq^2\chi_{0}\right]~. \label{RPAresp}
\eea
In this expression $\romk_F$ is the Fermi momentum while $m^*$
denotes the nucleon effective mass. The functions $\chi_0$, $\chi_2$
and $\chi_4$ are generalized free response functions, defined
as~\cite{gar92}
\be \chi_{2i} = \frac{1}{2}\langle \left[\left(\frac{{\romk}^2}
{\romk_F^2}\right)^i + \left(\frac{\vert \bf k + \bf q \vert
^2}{\romk_F^2}\right)^i\right]G_{HF} \rangle~, \ee
with $\chi_0=\langle G_{HF} \rangle$, and $\chi_{HF}=g\chi_0$.

The coupling between the two spin channels appears implicitly in the
function $\widetilde{W}_1^{(\alpha)}$ which can be expressed in
terms of the quantities $\beta_i$ introduced in Ref.\cite{gar92}
(their definitions are recalled in the appendix). One obtains:
\be \widetilde{W}_1^{(\alpha)} = W_1^{(\alpha)} +
C^{(\alpha)}w^2(I)W_{so}^2 \, \romq^4 \,
\frac{\beta_2-\beta_3}{1+W_2^{(\alpha')} \romq^2 (\beta_2-\beta_3)},
\label{new-w1} \ee
where $\alpha'$ is defined with respect to $\alpha$ as $S'=1-S$,
$I'=I$ (the third components $M^\prime$ and $Q^\prime$ are
irrelevant here since $W^{(\alpha')}_2$ does not depend on them),
and
\bea C^{(\alpha)} & = & 1 \quad \mathrm{if} \quad S=0 \nonumber
\\ & = & \frac{1}{2}M^2 \quad \mathrm{if} \quad S=1 \eea
If we replace $\widetilde{W}_1^{(\alpha)}$ in Eq.(\ref{RPAresp}) by
$W_1^{(\alpha)}$ we obtain the results of Ref.\cite{gar92}, as it
should be. It is worth noticing that the spin-orbit interaction
induces a complex coupling between the $S$=0 and $S$=1 channels.
This coupling is seen in Eqs.(A1-A3) of the appendix. The
$\beta_i$'s are complex and therefore, $\widetilde{W}_1^{(\alpha)}$
is a complex function of $\romq$ and $\omega$.

Finally, the quantity of interest is the dynamical structure
function $S^{(\alpha)}(\romq,\omega)$. At zero temperature, it is
just proportional to the imaginary part of the response function at
positive energies. At finite temperature T, one can also relate it
to the response function by the detailed balance relation and
obtain\cite{vau96}:
\be S^{(\alpha)}(\romq,\omega,T) = -\frac{1}{\pi} \frac{\rm{Im}
\chi^{(\alpha)}(\romq,\omega,T)} {1-e^{-\omega/T}}~, \ee
where the energy $\omega$ can be either positive or negative.

\section{Results}

We apply the above formalism to the case of neutron matter. The
calculations are performed with the Skyrme interaction SLy230b which
is designed for reproducing the neutron matter equation of
state\cite{cha97}. We can make a global assessment of our numerical
accuracy by calculating the energy-weighted integrals of our
dynamical structure functions, and compare them with the
energy-weighted sum rule obtained by the double commutator
method~\cite{boh79}. Indeed, the sum rule value must not depend on
the spin-orbit interaction.

\subsection{The spin-orbit induced interaction}
The function $\widetilde{W}_1^{(\alpha)}$ describes the coupling
between spin channels induced by the spin-orbit interaction. While
the interaction parameter $W_1^{(\alpha)}$ is real and independent
of $\omega$ and of the temperature, the new function
$\widetilde{W}_1^{(\alpha)}$ is complex and it depends on both
$\omega$ and $T$ through the $\beta_2$, $\beta_3$ terms.

\begin{figure}[h!]
\center
\includegraphics[scale=0.5,clip]{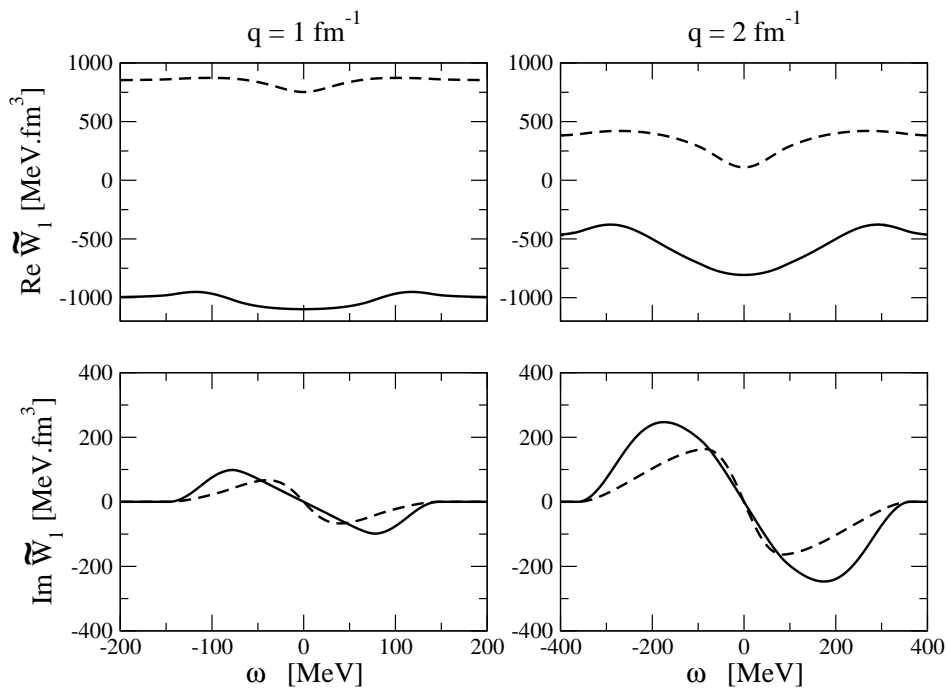}
\caption{The interaction parameter
$\widetilde{W}_1^{(\alpha)}(\romq,\omega)$ as a function of
$\omega$, for two different values of the momentum transfer. Solid
and dashed lines correspond to $S=0$ and $S=1$ channels,
respectively. The results are for neutron matter at $T=0$ and
$\rho$=$\rho_0$.} \label{figa1-1}
\end{figure}

In Fig.~\ref{figa1-1} are plotted the real and imaginary parts of
$\widetilde{W}_1^{(\alpha)}(\romq,\omega,T)$, as a function of
$\omega$ for two different values of $\romq$. The density has been
fixed to the value $\rho_0$ of saturation density of symmetric
nuclear matter, and the temperature is $T=0$. It is interesting to
notice that the $\omega$-dependence of the difference $\beta_2-
\beta_3$ has the same symmetry properties as the response function,
namely, the real (imaginary) part is symmetric (antisymmetric) with
respect to $\omega$=0 for a fixed value of $\romq$. This property is
fulfilled in spite of the fact that each of the functions $\beta_2$
and $\beta_3$ does not satisfy it separately. Such symmetry
properties are seen on Fig.~\ref{figa1-1}.

When $\omega \to \pm \infty$, both $\beta_i$ functions of
Eq.(\ref{new-w1}) go to zero and $\widetilde{W}_1^{(\alpha)}$ tends
to ${W}_1^{(\alpha)}$. Therefore, the amplitude of the oscillations
of the curves in Fig.~\ref{figa1-1} show the deviations of
$\widetilde{W}_1^{(\alpha)}$ with respect to ${W}_1^{(\alpha)}$. As
for the real parts, these deviations are most visible around
$\omega$=0. For the imaginary parts, they are always zero at
$\omega$=0 and the deviations become significant at larger values of
$\vert \omega \vert$. In any case, one can see that the change of
the p-h interaction due to the spin-orbit force seems to be
relatively small, except at higher values of the transferred
momentum $\romq$. This increase with $\romq$ reflects the $\romq^4$
power explicitly appearing in Eq.(\ref{new-w1}).

\subsection{Dynamical structure functions}

\begin{figure}[h!]
\center
\includegraphics[scale=0.5,clip]{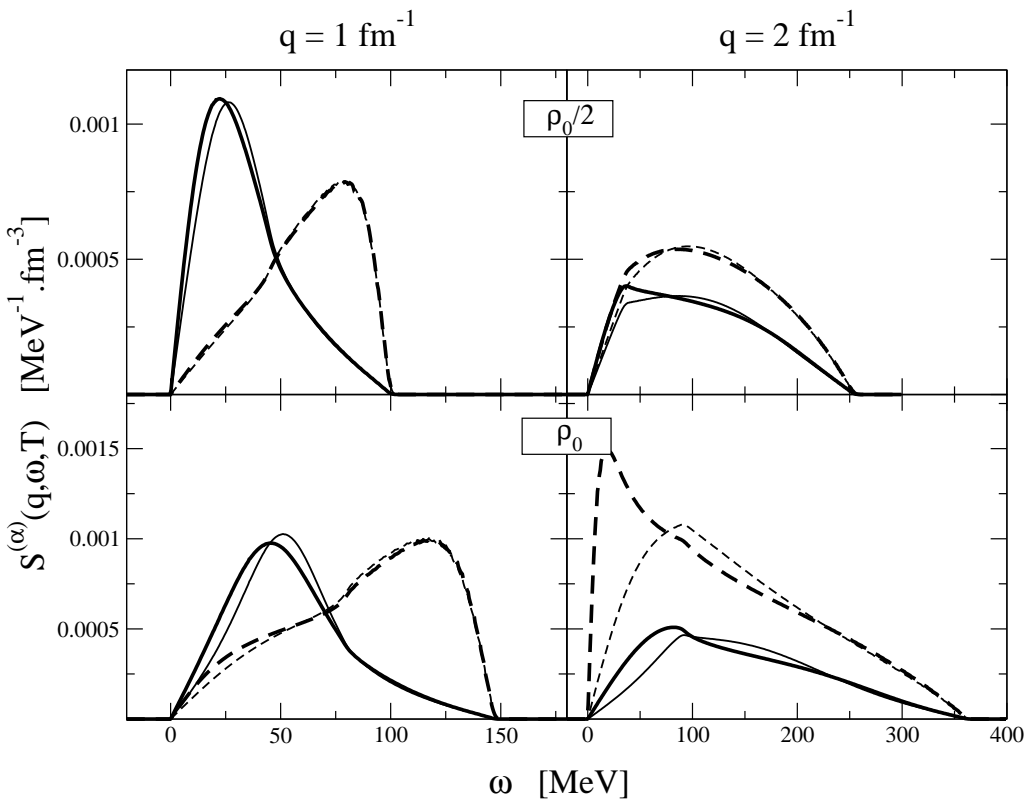}
\caption{The dynamical structure function
$S^{(\alpha)}(\romq,\omega)$ at $T$=0 as a function of $\omega$, for
two values of the momentum transfer $\romq$ and at densities $\rho =
\rho_0/2$ and $\rho_0$ . Solid and dashed lines correspond to $S=0$
and $S=1$ channels, respectively. The thin lines represent the
structure function without spin-orbit interaction.} \label{figa1}
\end{figure}

\begin{figure}[h!]
\center
\includegraphics[scale=0.5,clip]{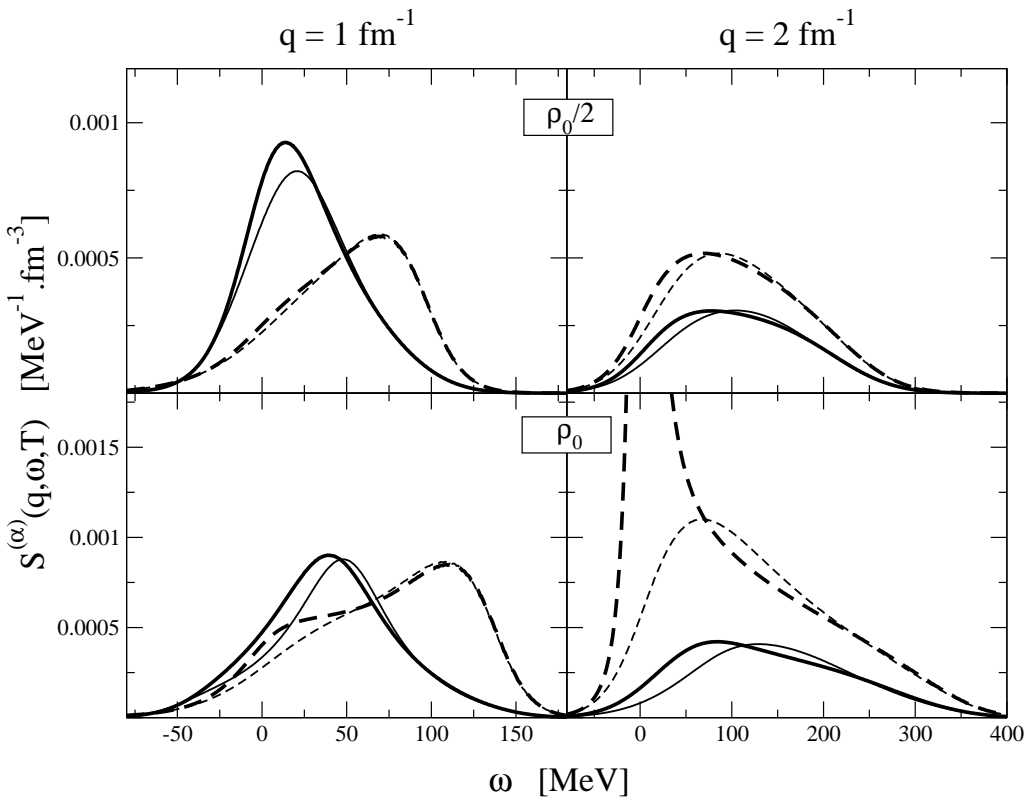}
\caption{Same as Fig.\ref{figa1}, for $T=$20 MeV.} \label{figa2}
\end{figure}

We now turn to the effect of the spin-orbit interaction on the
dynamical structure function $S^{(\alpha)}(\romq,\omega)$. In
Figs.~\ref{figa1}-\ref{figa2} are plotted the values of
$S^{(\alpha)}(\romq,\omega)$ as functions of $\omega$, calculated at
zero temperature and at $T$=20 MeV. The calculations are done at the
fixed values of momentum transfer $\romq$ and neutron matter
densities indicated in the figures. We show separately the $S$=0 and
$S$=1 cases, calculated with and without the spin-orbit force. A
general observation is that the effect of spin-orbit interaction
increases as $\romq$ increases. This can be easily understood
according to the previous analysis of the interaction parameter
$\widetilde{W}_1^{(\alpha)}(q,\omega)$.

We first discuss the $T$=0 results of Fig.~\ref{figa1}. For the
smaller value of $\romq$ the effect increases slightly with
increasing density, this effect being more visible in the $S$=0
channel. For the larger value of $\romq$ the increasing effect with
increasing density becomes more dramatic in the $S$=1 channel. The
reason for this behaviour is due to the fact that, at $\rho=\rho_0$
and $q$=2 fm$^{-1}$ we are approaching the point of instability in
the $S$=1 channel. This instability is characteristic of Skyrme-type
interactions~\cite{mar02}. Indeed, the response function depends not
only on $\widetilde{W}_1^{(\alpha)}$ but also on $W_2^{(\alpha)}$ as
shown in Eq.(\ref{RPAresp}). Now, the interaction SLy230b gives
$W_2^{S=0}$=163.55 MeV.fm$^5$, $W_2^{S=1}=-$163.55 MeV.fm$^5$. From
Fig.~\ref{figa1-1} one also sees that the spin-orbit force
contributes some important negative amount in the $S$=1 channel,
thus enhancing the tendency to instability. One can conclude that
the effect of the spin-orbit force is generally small but it can
become dramatic when one approaches the instability region of the
$S$=1 channel.

The above conclusions remain valid at finite temperature but they
are further amplified, as one can see from Fig.~\ref{figa2} in the
case of $T$=20 MeV. As expected, noticeable modifications of the
dynamical structure function are found for $\omega$ around zero and
$\romq$ larger than $\romk_F$. One must keep in mind that the
effective interaction parameter
$\widetilde{W}_1^{(\alpha)}(\romq,\omega)$ containing the
contribution of the spin-orbit force is temperature dependent
through the $\beta_2$, $\beta_3$ functions. In the $T$=20 MeV case
the vicinity of the instability point is clearly seen for
$\rho=\rho_0$ and $q$=2~fm$^{-1}$.

\subsection{The neutrino mean free paths }

We examine now the effect of the spin-orbit interaction on neutrino
mean free paths in neutron matter under various conditions of
density and temperature. The scattering of neutrinos on neutrons is
mediated by the neutral current of the electro-weak interaction. In
the non-relativistic limit and in the case of non-degenerate
neutrinos, the mean free path $\lambda$ of a neutrino with initial
momentum ${\bf k}_1$ is given by~\cite{iwa,reddy}
\be 1/ \lambda({\rm k}_1,T) = \frac{G_F^2}{32 \pi^3} \int d{\bf k}_3
\left( c_V^2 (1+\cos{\theta})~{\cal S}^{(0)}(q,T)
 + c_A^2 (3-\cos{\theta})~{\cal S}^{(1)}(q,T) \right)~,
\ee
where $T$ is the temperature, $G_F$ is the Fermi constant, $c_V$
($c_A$) the vector (axial) coupling constant. The final neutrino
momentum is ${\bf k}_3$, the four-vector $q = k_1-k_3$ stands for
the transferred energy-momentum, and $\cos\theta=\hat{\bf
k}_1\cdot\hat{\bf k}_3$. The dynamical structure factors ${\cal
S}^{(S)}(q,T)$ describe the response of neutron matter to
excitations induced by neutrinos, and they contain the relevant
information on the medium. The vector (axial) part of the neutral
current gives rise to density (spin-density) fluctuations,
corresponding to the $S=0$ ($S=1$) spin channel.

We have calculated the neutrino mean free path at different
densities ($\rho_0$ and $2\rho_0$) and temperatures (10, 20 and 30
MeV). The energy of the incoming neutrino is chosen to be
$E_\nu=3T$. Results are shown in Table~\ref{table2}. The first line
($\lambda_{HF}$) shows the results for the neutrino mean free path
calculated at the self-consistent mean field approximation, i.e.,
without RPA correlation effects. The spin-orbit interaction has no
effect in this case, since we are in a homogeneous medium and the
spin-orbit force does not contribute to the HF properties. Next, we
show the results of the complete calculation with or without
spin-orbit interaction. For $T=10$ MeV, the spin-orbit interaction
modify the mean free path by only 1\%, 3-4\% for $T=20$ MeV and
5-10\% for $T=30$ MeV. We thus conclude that the effects of the
spin-orbit interaction on the neutrino mean free paths are at the
level of a few percent.

\begin{table}[h]
\centering
\begin{tabular}{c|c|c|c|c|c|c|c|}
\cline{2-8}
& $\rho/\rho_0$ & \multicolumn{3}{c|}{1} & \multicolumn{3}{c|}{2} \\
\cline{2-8}
& T (MeV) & 10 & 20 & 30 & 10 & 20 & 30 \\
\cline{2-8} \hline
\multicolumn{1}{|c|}{$\lambda_{HF}$ [m]} & - & 45.1 & 5.83 & 1.84 & 67.0 & 8.19 & 2.33  \\
\hline
\multicolumn{1}{|c|}{$\lambda_{RPA}/\lambda_{HF}$} & with spin-orbit & 1.03 & 0.93 & 0.79 & 0.54 & 0.44 & 0.27 \\
\cline{2-8} \multicolumn{1}{|c|}{~}                      & without
spin-orbit & 1.04 & 0.96 & 0.83 & 0.55 & 0.48 & 0.38 \\ \hline
\end{tabular}
\caption{Neutrino mean free paths (in meters) in neutron matter,
calculated in HF and RPA schemes. The calculations correspond to the
different densities and temperatures as indicated. The neutrino
energies are $E_\nu=3T$. } \label{table2}
\end{table}

\section{Concluding remarks}
We have investigated the effects of the spin-orbit component of the
p-h interaction $V_{ph}$ on the RPA nuclear response functions and
their possible consequences on the neutrino mean free paths. This
study is carried out in the framework of a Skyrme-type, zero-range
effective interaction. While the central component of $V_{ph}$ keeps
the $S$=0 and $S$=1 spin channels separated, the spin-orbit
component couples these channels together. However, within the
specific form of Skyrme-type interactions this coupling appears only
implicitly through a modified interaction parameter
$\widetilde{W}_1^{(\alpha)}(\romq,\omega,T)$, and the calculation of
the response function is formally identical to the case without
spin-orbit interaction.

The modified interaction parameter $\widetilde{W}_1^{(\alpha)}$ is
shown to be complex and it depends on the energy-momentum transfer
($\omega,\romq$) and temperature $T$. Its behaviour at large
$\omega$ shows that the effect of the spin-orbit force tends to zero
for increasing $\omega$. The overall effects on the response
functions remain small in neutron matter at densities up to
$\rho_0$. However, in the specific example of the SLy230b force that
we have considered, a pole in the response function and hence an
instability occur in the $S$=1 channel at $\rho \simeq \rho_0$ and
$\romq \simeq 2$ fm$^{-1}$. In this case, even a small modification
brought about by the spin-orbit force produces a large change of the
$S$=1 response function near the pole.

As for the $T$-dependence of the spin-orbit effects, all the remarks
made above remain true with increasing $T$, the only difference
being that the effects are amplified at higher temperature. Finally,
the neutrino mean free paths in neutron matter are very moderately
affected by the spin-orbit component of the p-h interaction.

The numerical applications have been presented here for the case of
neutron matter, but similar results are obtained in symmetric and
asymmetric nuclear matter.We also note that the zero-range nature of
the spin-orbit force studied here is reflected in the $\romq^4$
dependence of the modified p-h interaction, and therefore the fact
that the spin-orbit effects increase with increasing $q$ would be
altered for a finite range interaction.

\subsection*{Acknowledgments}

This work is supported in part by the grant FIS2004-0912 (MEC,
Spain), the IN2P3(France)-CICYT(Spain) exchange program and the
Financement Th\'eorie IN2P3 fund.

\appendix

\section{The coupled integral equations for the $\grpa$ Green's functions}
In this Appendix we shall show the way to transform the
Bethe-Salpeter equation (\ref{eqBS}) into a set of coupled algebraic
equations for three integrated quantities depending on
$G^{(\alpha)}_{RPA}$. To alleviate the notation, we will omit the
$(\romq,\omega)$-dependence, and specify only the spin variables, as
there is no isospin coupling. The expressions are formally valid for
both symmetric nuclear matter and neutron matter. The only
differences will be in the value of the spin-isospin degeneracy
factor $g$ of Eq.~(\ref{chi}), and the factor $w(I)$ of
Eq.~(\ref{vph-so}).

 One must note that the multipole
expansion of $G_{HF}$ (see Eq.~\ref{GHF}) only involves terms of the
type $Y_{L,0}(\hat{k}_1)$. Therefore integrals of the type $\langle
f(\romk) Y_{L,M} G_{HF} \rangle$ or $\langle f(\romk) Y_{L,M}
Y_{L',M'} G_{HF} \rangle$ vanish unless $M \ne 0$ or $M+M' \ne 0$,
respectively. This will simplify the response function equations.
Other integrals involving $G_{HF}$ are also needed and they can be
expressed in terms of the quantities $\beta_i$ introduced in
\cite{gar92}: \be
\begin{array}{lll}
\langle \gz \rangle = \beta_0~, & \quad \langle \romk^2 \gz \rangle
= \romq^2 \beta_2~, & \langle \romk^4 \gz \rangle = \romq^4
\beta_5~,
\\ [1mm]
\langle \romk Y_{1,0} \gz \rangle = \romq \sqrt{\frac{3}{4 \pi}}
\beta_1~, & \quad \langle \romk^3 Y_{1,0} \gz \rangle = \romq^3
\sqrt{\frac{3}{4 \pi}} \beta_4~, &
\\ [1mm]
\langle \romk^2 |Y_{1,0}|^2 \gz \rangle = \romq^2 \frac{3}{4 \pi}
\beta_3~, & \quad \langle \romk^2 |Y_{1,1}|^2 \gz \rangle = \romq^2
\frac{3}{8 \pi} \left( \beta_2 - \beta_3 \right)~. &
\end{array}
\nonumber
\ee

Let first consider the $S=0$ channel. With the p-h interaction given by
Eqs.~(\ref{vph-no}-\ref{vph-so}), the Bethe-Salpeter equation is written as
\bea
\grpaz({\bf k}_1)
&=& G_{HF}({\bf k}_1) + W_1^{(0)} G_{HF}({\bf k}_1) \langle \grpaz \rangle
+ W_2^{(0)} \romk_1^2 G_{HF}({\bf k}_1) \langle \grpaz \rangle \nonumber \\
&& + W_2^{(0)} G_{HF}({\bf k}_1) \langle \romk^2 \grpaz \rangle - 2
W_2^{(0)} \frac{4 \pi}{3} \sum_{\mu} \romk_1 Y^*_{1 \mu}(\hat{1})
G_{HF}({\bf k}_1) \langle \romk Y_{1 \mu} \grpaz \rangle \nonumber \\
&& - 2 \sqrt{\frac{4 \pi}{3}} W_{so} \, \romq \, \sum_{M_S^\prime} M_S^\prime \,
\romk_1 Y_{1M_S^\prime}(\hat{1}) G_{HF}({\bf k}_1) \langle \grpaump \rangle \nonumber \\
&& + 2 \sqrt{\frac{4 \pi}{3}} W_{so} \, \romq \, \sum_{M_S^\prime}
M_S^\prime \, G_{HF}({\bf k}_1) \langle \romk
Y_{1M_S^\prime}\grpaump \rangle~. \label{eq:S0} \eea Integrating
over $\bf{k}_1$ we get \bea \langle \grpaz \rangle &=& \beta_0 +
W_1^{(0)} \beta_0 \langle \grpaz \rangle + W_2^{(0)} \romq^2 \beta_2
\langle \grpaz \rangle
 + W_2^{(0)} \beta_0 \langle \romk^2 \grpaz \rangle -
2 W_2^{(0)} \sqrt{\frac{4 \pi}{3}} \romq \beta_1
\langle \romk Y_{1 0} \grpaz \rangle \nonumber \\
&& + 2 \sqrt{\frac{4 \pi}{3}} W_{so} \, \romq \, \beta_0
\sum_{M_S^\prime} M_S^\prime \, \langle \romk
Y_{1M_S^\prime}\grpaump \rangle~. \label{eq:S0int} \eea One can see
that the quantity $\langle \grpaz \rangle$ we are interested in is
coupled to $\langle \romk^2 \grpaz \rangle$, $\langle \romk Y_{10}
\grpaz \rangle$, and $\langle \romk Y_{1M_S^\prime}\grpaump
\rangle$. Two new equations are obtained multiplying
Eq.~(\ref{eq:S0}) with $\romk_1^2$ and $\romk_1 Y_{10}(\hat{k}_1)$,
and integrating over $\bf{k}_1$. These factors are such that there
is no contribution from the term $\langle \grpaump \rangle$ entering
Eq.~(\ref{eq:S0}). The coupling with the $S=1$ channel is thus
contained in the last term entering Eq.~(\ref{eq:S0int}). From the
Bethe-Salpeter equation for the $S=1$ channel the following
expression is obtained \be \sum_{M_S^\prime} \langle M_S^\prime
\romk Y_{1,M_S^\prime} \grpaum \rangle = 2 \, \sqrt{\frac{3}{4 \pi}}
W_{so} \romq^3 \frac{\beta_2-\beta_3}{1+W_2^{(1)} \romq^2
(\beta_2-\beta_3)} \langle \grpaz \rangle~. \label{eq:subs1} \ee
This means that the effect of the spin-coupling can be simply
absorbed in an effective $\widetilde{W}_1^{(0)}$ coefficient. The
equations for $S=1$ channel are obtained proceeding along similar
lines. The explicit expression of $\widetilde{W}_1^{(\alpha)}$ is
given in Eq.(\ref{new-w1}).

Finally, the system of algebraic equations can be written in a
compact form for both channels as \bea \left( 1 -
\widetilde{W}_1^{(\alpha)} \beta_0 - W_2^{(\alpha)} \romq^2 \beta_2
\right) \langle G_{RPA}^{(\alpha)} \rangle - W_2^{(\alpha)} \beta_0
\langle \romk^2 G_{RPA}^{(\alpha)} \rangle + 2 W_2^{(\alpha)} \romq
\beta_1 \sqrt{\frac{4 \pi}{3}} \langle \romk Y_{10}(\hat{k})
G_{RPA}^{(\alpha)} \rangle
&=& \beta_0 \nonumber \\
- \left(\widetilde{W}_1^{(\alpha)} \romq^2 \beta_2 + W_2^{(\alpha)}
\romq^4 \beta_5 \right) \langle G_{RPA}^{(\alpha)} \rangle +
\left(1- W_2^{(\alpha)} \romq^2 \beta_2 \right) \langle \romk^2
G_{RPA}^{(\alpha)} \rangle + 2 W_2^{(\alpha)} \romq^3 \beta_4
\sqrt{\frac{4 \pi}{3}} \langle \romk Y_{10}(\hat{k})
G_{RPA}^{(\alpha)} \rangle
&=& \romq^2 \beta_2 \nonumber \\
- \left(\widetilde{W}_1^{(\alpha)} \romq \beta_1 + W_2^{(\alpha)}
\romq^3 \beta_4 \right) \langle G_{RPA}^{(\alpha)} \rangle -
W_2^{(\alpha)} \romq \beta_1 \langle \romk^2 G_{RPA}^{(\alpha)}
\rangle + \left(1+2 W_2^{(\alpha)} \romq^2 \beta_3 \right)
\sqrt{\frac{4 \pi}{3}} \langle \romk Y_{10}(\hat{k})
G_{RPA}^{(\alpha)} \rangle &=& \romq \beta_1 \nonumber \eea

\end{document}